\documentclass[conference]{IEEEtran}
\IEEEoverridecommandlockouts

\newcommand{\E}[0]{\mathbb{E}}

\usepackage{cite}
\usepackage{amsmath,amssymb,amsfonts}
\usepackage{algorithmic}
\usepackage{graphicx}
\usepackage{textcomp}
\usepackage{url}
\usepackage{xcolor}

\begin{document}

\title{Resampled Mutual Information for Clustering and Community Detection}

\author{\IEEEauthorblockN{Cheaheon Lim}
\IEEEauthorblockA{\textit{Harvard University} \\
Cambridge, MA, USA \\
Email: ilim@college.harvard.edu}
}

\maketitle

\begin{abstract}
We introduce resampled mutual information (ResMI), a novel measure of clustering similarity that combines insights from information theoretic and pair counting approaches to clustering and community detection. Similar to chance-corrected measures, ResMI satisfies the constant baseline property, but it has the advantages of not requiring adjustment terms and being fully interpretable in the language of information theory. Experiments on synthetic datasets demonstrate that ResMI is robust to common biases exhibited by existing measures, particularly in settings with high cluster counts and asymmetric cluster distributions. Additionally, we show that ResMI identifies meaningful community structures in two real contact tracing networks.
\end{abstract}


\section{Introduction}

Clustering and community detection are essential tools in the analysis of complex systems, including social networks, biological systems, and document corpora. Evaluation metrics for clustering results generally fall into three categories: information theoretic measures, such as normalized  mutual information (NMI) and its variants \cite{vinh_first_nmi, vinh_review, newman_rmi, graph_simulation_study}, pair counting measures, such as the Rand Index (RI) and its variants \cite{rand_original, ARI_original, ARI_AMI_issue, ARI_issue}, and set matching measures, such as the variation of information \cite{MEILA2007873}. Here, we focus on unifying and addressing the limitations of the former two classes of measures. 

Intuition suggests that a measure of clustering similarity should satisfy the constant baseline property: the expected similarity between two independently drawn clusterings  equals zero. However, neither the NMI nor RI meet this criterion, and there exists a rich literature on addressing this issue \cite{vinh_first_nmi, vinh_review, newman_rmi, graph_simulation_study, ARI_original, ARI_AMI_issue, ARI_issue}. Of the possible solutions, modifications based on chance correction, which can be applied to both NMI and RI, are the most well-known \cite{vinh_first_nmi, ARI_original}. The central idea behind chance correction is to adjust each measure by the expected similarity between two independently sampled clusterings via the permutation model, where cluster divisions are randomly drawn while holding  the number and sizes of clusters constant. 

Although measures adjusted in this manner satisfy the constant baseline property, they come at the cost of interpretability and are no longer constrained to the unit interval. Moreover, recent work in \cite{ARI_AMI_issue} highlights that the assumptions of the permutation model often fail to hold in practice, and modifying the adjustment term to more accurately reflect the clustering algorithm being implemented can lead to drastically different conclusions. For instance, repeated K-means clustering yields a fixed number of clusters, but the distribution of cluster sizes can vary significantly. Correctly specifying the model of randomization and chance correction term results in the rankings of clusterings reversing in certain settings, which illustrates the potential drawbacks of applying the chance-corrected versions of NMI and RI that rely on the permutation model. This finding is also concerning in the context of assessing the performance of distinct clustering algorithms, as correctly specifying the chance correction term for each procedure will result in different similarity measures.

For this reason, we argue that a similarity measure for comparing clusterings must satisfy model independence in addition to the constant baseline property. In the following section, we propose resampled mutual information (ResMI), a novel measure of clustering similarity that applies insights from both information theoretic and pair counting approaches to clustering and community detection. Despite its relatively simple formulation, ResMI satisfies the model independence and constant baseline properties, and experiments on synthetic datasets show that ResMI is robust against common biases of existing clustering measures that have been documented in the literature \cite{vinh_first_nmi, vinh_review, newman_rmi, ARI_issue}. We also assess  the performance of ResMI against other information theoretic and pair counting measures of clustering similarity using the contact tracing networks of \cite{contact_data_1}, \cite{contact_data_2}, and  find that ResMI consistently identifies meaningful community structures.

\section{Measures of Clustering Similarity}

\subsection{NMI and Variants}

Consider a set of objects $[n]:= \{1,...,n\}$ and a labeling procedure $f: [n] \to [M]$ that assigns each object $i \in [n]$ some label $m \in [M]$. Define $m_f:= || \{ i \in [n] : f(i) = m\}||$ as the number of objects assigned label $m$ by procedure $f$. When making comparisons between two different procedures $f: [n] \to [M]$ and $g: [n] \to [M']$, we take $m_f \wedge m'_g := || \{ i \in [n] : f(i) = m, g(i) = m'\}||$ to be the number of objects assigned label $m$ by $f$ and $m'$ by $g$. In the literature, the following expression is referred to as the mutual information between the labelings $f$ and $g$,
\begin{align}
    I(P_f;P_g) = \sum_{m, m'} P_{f,g}(m,m') \log \frac{P_{f,g}(m,m')}{P_f(m) P_g(m')}, \label{eq_MI}
\end{align}
where $P_{f,g}(m, m') := \frac{m_f \wedge m'_{g}}{n}$, $P_f(m) := \frac{m_f}{n}$, and $P_g(m') := \frac{m_g'}{n}$. To constrain the measure to the unit interval, a normalized version was proposed in \cite{NMI_1, NMI_3}, defined as
\begin{align}
    \text{NMI}(f; g) := \frac{I(P_f; P_g)}{\frac{1}{2} (H(P_f) + H(P_g) )},
\end{align}
where $H(P_f) = - \sum_{m} P_f(m) \log P_f(m)$. Other normalization schemes, such as dividing \eqref{eq_MI} by $\max\{ H(P_f) , H(P_g)\}$ or $\min\{ H(P_f) , H(P_g)\}$, are surveyed in \cite{vinh_review}. 

One important limitation of similarity measures based on  \eqref{eq_MI}  is their bias for higher cluster numbers \cite{NMI_bias_proof}. As an extreme example, consider the trivial labeling $f: [n] \to [n]$ that assigns each object a unique label. In this setting, any reasonable measure of similarity between the labelings $f$ and $g$ ought to equal zero, as the former provides no information about the latter. However, because $P_{f,g}(m,m') \in \{0, \frac{1}{n}\}$ and  $P_f(m) = \frac{1}{n}$, the mutual information between $f$ and $g$ is
\begin{align}
    I(P_f;P_g) &  =  \sum_{m, m'} P_{f,g}(m,m') \log \frac{ 1_{\{ \exists i : f(i) =m , g(i) = m'\} } }{  P_g(m')}  \nonumber \\
    & =  \sum_{m'}   P_{g}(m') \log \frac{1}{P_g(m')}, \nonumber 
\end{align}
which equals $H(P_g)$, its \emph{maximum} possible value (where we take $0 \log 0 = 0$). The key issue behind this counterintuitive result is that the densities $P_f, P_g$ correspond to the empirical frequencies of the row and column marginals of  Table \ref{table_contingency}, whereas the true information content that labeling $f$ has about labeling $g$ is reflected by the entire contingency table.

\begin{table}[htbp]
\caption{Contingency Table Between Labelings}
\vspace{-27pt}
\begin{center}
$$ \begin{array}{c|cccc|c}
\mathbf{f} \backslash \mathbf{g} & 1 & 2 & \ldots & M' & \text { Sums } \\
\hline 1 & 1_f \wedge 1_g & 1_f \wedge 2_g  & \ldots & 1_f \wedge M'_g & 1_f \\
2 & 2_f \wedge 1_g & 2_f \wedge 2_g & \ldots & 2_f \wedge M'_g & 2_f \\
\vdots & \vdots & \vdots & \ddots & \vdots & \vdots \\
M & M_f \wedge 1_g & M_f\wedge  2_g & \ldots & M_f \wedge M'_g & M_f \\
\hline \text { Sums } & 1_g & 2_g & \ldots & M_g' & n
\end{array} $$
\label{table_contingency}
\end{center}
\end{table}

NMI's failure to satisfy the constant baseline property can be viewed as a consequence of its bias for labelings with a higher number of clusters.  A number of modifications to NMI have been proposed to address this issue, and the most well-known is the adjusted mutual information (AMI) of \cite{vinh_first_nmi}. AMI takes a chance correction approach by subtracting NMI by $\E[ \text{NMI}(f, g)]$, the expected NMI between two independent clusterings generated by randomly sampling from the row and column marginals of Table \ref{table_contingency} under the permutation model. Formally, AMI is defined as
\begin{align}
    \text{AMI}(f, g) : = \frac{\text{NMI}(f, g) - \E[ \text{NMI}(f, g)] }{1 - \E[ \text{NMI}(f, g)]}.
\end{align}

The reduced mutual information (RMI) of \cite{newman_rmi} considers the problem from an operational point of view, providing an upper bound on the mutual information between two labelings by approximating the number of bits a sender can save when transmitting Table \ref{table_contingency} if the receiver has knowledge about one of the row or column marginals. Formally, RMI is defined as
\begin{align}
    \text{RMI} := I(f;g) - \frac{1}{n} \log \Omega(f;g),
\end{align}
where $\Omega(f;g)$ is the  number of $M \times M'$ non-negative integer matrices with row and column marginals that coincide with Table \ref{table_contingency}. In practice, a normalized version of RMI is  used to allow for comparisons across different settings.

Although AMI and RMI address the cluster count bias of NMI and satisfy the constant baseline property, both  measures are no longer constrained to the unit interval and can take on negative values. As mentioned in the Introduction, AMI and chance-corrected methods in general also suffer from model dependence, as some randomization scheme must be assumed to specify the term $\E[ \text{NMI}(f, g)] $. The implementation of RMI may also be challenging in practice, as calculating $\Omega(f;g)$ is computationally expensive in large data settings. More generally, AMI and RMI come at the cost of interpretability and are not entirely satisfactory from an information theoretic standpoint. Both measures apply transformations to \eqref{eq_MI}, and they lack the clear interpretation mutual information has as a measure of shared information between two random variables. 

\subsection{Pair Counting Approaches} 

While there are a vast number of pair counting measures for comparing clusterings, we follow \cite{vinh_review} and focus on the Rand Index (RI) and its variants \cite{rand_original, ARI_original, ARI_issue, ARI_AMI_issue} that are most widely used in the literature. Continuing with the notation from the previous section, the RI between the labelings $f$, $g$ is defined as
\begin{align}
    \text{RI}(f,g) &: = \frac{ \sum_{i> j} 1_{ \{ f(i) = f(j), \hspace{1pt} g(i) = g(j) \} \vee  \{ f(i) \neq f(j), \hspace{1pt} g(i) \neq g(j) \} }  }{ \binom{n}{2} } , 
\end{align}
interpreted as the proportion of pairs of objects in $[n]$ that have the same classification result (assigned to the same or different cluster) under $f$ and $g$. Because the expected number of equivalently-classified pairs is non-zero even for two independently generated clusterings, RI does not satisfy the constant baseline property. As a result, the Adjusted Rand Index (ARI) is used more frequently in practice, which is the chance-corrected version of RI that adjusts for the expected number of chance agreements between $f$ and $g$ in the exact manner that AMI adjusts NMI. Formally, ARI is defined as
\begin{align}
    \text{ARI} & := \frac{\text{RI} - \mathbb{E}[\text{RI}]}{1 - \mathbb{E}[\text{RI}]} ,
\end{align}
where $\mathbb{E}[\text{RI}]$ is the expected RI when clusterings are randomly drawn from the row and column marginals of Table \ref{table_contingency} under the permutation model.

As was the case with AMI, ARI is no longer constrained to the unit interval and can take on negative values, and the chance-correction comes at the cost of interpretability and model independence, as specifying  the term $\E[\text{RI}]$ requires the assumption of some randomization scheme \cite{ARI_AMI_issue}. More generally, while RI-based measures do not suffer from the cluster count bias of NMI, they have a demonstrated bias for symmetric cluster distributions, tending to misrepresent the similarity of clusterings when cluster sizes vary widely \cite{ARI_issue}. 

\subsection{Resampled Mutual Information (ResMI)}

ResMI unifies the information theoretic and pair counting approaches to comparing clusterings by fundamentally shifting the focus of equation \eqref{eq_MI}, instead considering the mutual information between cluster labelings as captured by a pair of random samples. Formally, let $Z_1, Z_2$ be a pair of random objects drawn \emph{without} replacement from $[n]$. ResMI is the  normalized  mutual information between the indicator variables for the labels of $Z_1$ and $Z_2$ being equivalent under $f$ and $g$, defined as
\begin{align}
    \text{ResMI}(f,g) & :=  \frac{I( 1_{ \{ f(Z_1) = f(Z_2) \} } ; 1_{ \{g(Z_1) = g(Z_2) \} } )  }{ \frac{1}{2} (H(1_{ \{ f(Z_1) = f(Z_2) \} } )+  H(1_{ \{ g(Z_1) = g(Z_2) \} } ) ) } .
\end{align}

Similar in spirit to pair counting measures of similarity, the ``randomness" in the clustering results are captured by $1_{ \{ f(Z_1) = f(Z_2) \} }$ and $1_{ \{g(Z_1) = g(Z_2) \} }$, two Bernoulli random variables  parameterized by  $q_f := (1/ \binom{n}{2}) \sum_{i> j} 1_{\{ f(i) = f(j) \}}$ and $q_g := (1/{\binom{n}{2}} ) \sum_{i > j}  1_{\{ g(i) = g(j) \}}$, respectively. The numerator  term, denoted $I_{\text{ResMI}}(f;g)$, can be calculated using the  probability that two objects have the same label under $f$, conditional on having the same label under $g$ (or vice versa). Let $\mathcal{G}:= \{ (i,j ) : g(i) = g(j), \hspace{3pt} i>j \}$ denote the set of pairs of objects that have the same label under $g$, and define the conditional probabilities $q_{f|\mathcal{G}}:=  (1/\binom{|\mathcal{G}|}{2})  \sum_{ (i,j) \in \mathcal{G} }  1_{ \{ f(i) = f(j) \} }$ and $q_{f|\mathcal{G}^c}:= (1/\binom{|\mathcal{G}^c|}{2}) \sum_{ (i,j) \in \mathcal{G}^c }  1_{ \{ f(i) = f(j) \} }$. Then, $I_{\text{ResMI}}(f;g)$ equals,
\begin{align}
      I_{\text{ResMI}}(f;g) &  = h_b(q_f) - \Big[  q_g  h_b( q_{f|\mathcal{G}} ) + (1-q_g) h_b( q_{f| \mathcal{G}^c} ) \Big]. \nonumber
\end{align}

It is clear that $I_{\text{ResMI}}(f;g)$ no longer suffers from the issue of cluster count bias when $f: [n] \to [n]$ is the trivial clustering, in which case $q_f = q_{f|\mathcal{G}} = q_{f|\mathcal{G}^c} = 0$ and $I_{\text{ResMI}}(f;g) = 0$, as no two objects have the same label under $f$. Experimental results in the subsequent section establish this result in generality, and we show that ResMI satisfies the constant baseline property without requiring chance correction terms in its formulation as other methods do. Note that mutual information as originally defined in \eqref{eq_MI} is equivalent to  $I( f(Z_1), g(Z_1) )$ in our new notation, illustrating how $I_{\text{ResMI}}(f;g)$ fundamentally departs from NMI-based approaches in its encoding of the randomness of clusterings. Finally, by being formulated completely in the language of information theory, ResMI retains its interpretability as the measure of information shared by two random variables associated with the labelings $f,g$.

\section{Experiments on Synthetic Data}

We perform a comparative analysis of ResMI against NMI, AMI, ARI, and RMI on four different synthetic datasets that simulate settings where existing clustering metrics have demonstrated limitations. The R library \texttt{aricode} \cite{aricode} is employed to implement NMI, AMI, and ARI, and the R library \texttt{clustAnalytics} is used to compute normalized RMI \cite{clustAnalytics}. 

\begin{figure}[htbp] 
\centering
\includegraphics[width=9cm]{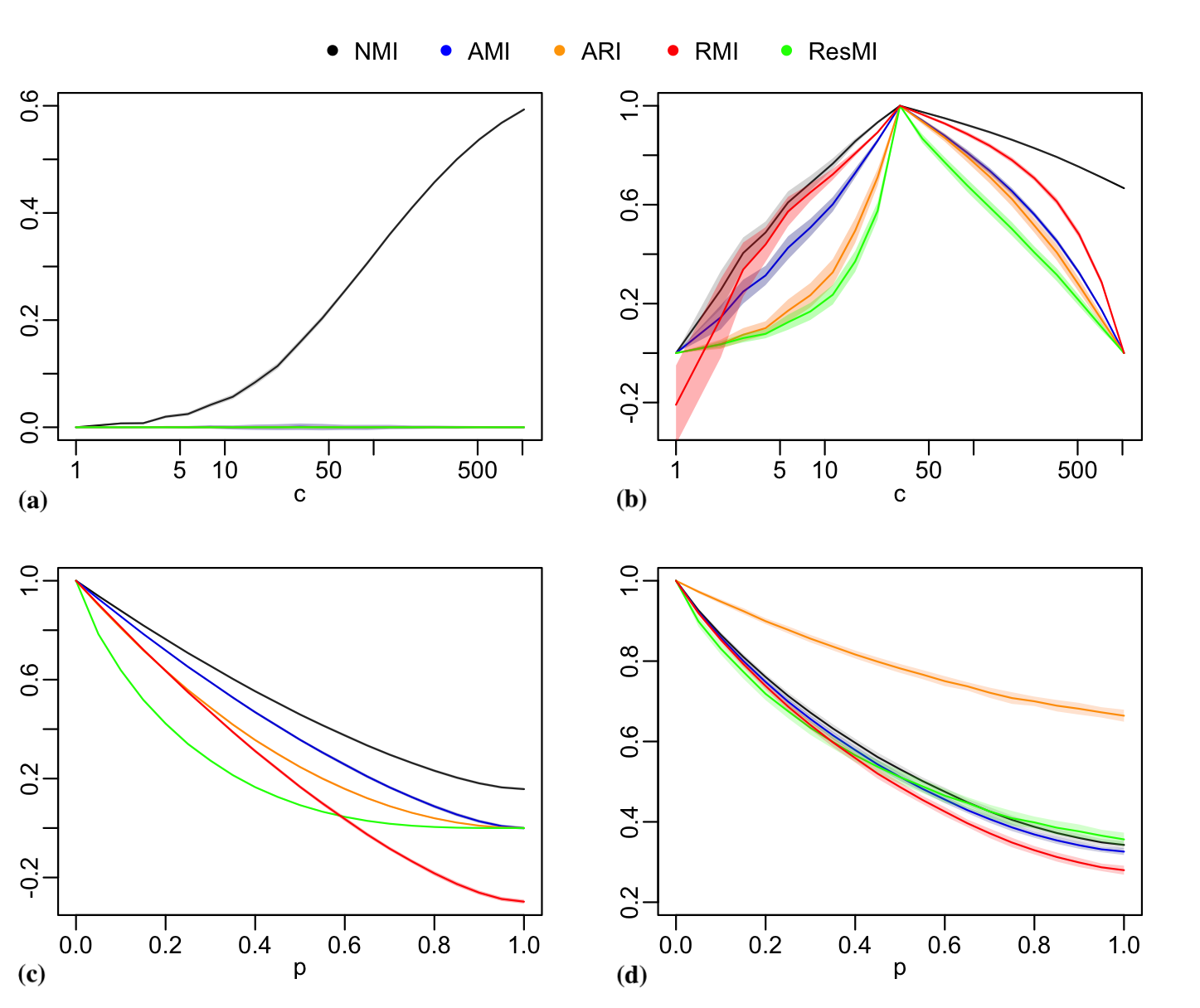}
\caption{Results from synthetic data experiments, averaged over 100 runs with error bars denoting one standard deviation. In (a)-(c), the ground truth consists of 1,024 objects in 32 equally sized clusters. Similarity with the ground truth is plotted for clusterings generated by (a) random assignment to $c$ clusters, (b) random merging/splitting of ground truth clusters  to reach  $c$ clusters, and (c) random shuffling of $p$ proportion of labels.  In (d), the ground truth has 512 objects in one cluster, 256 in another, and the remainder evenly distributed across 5 others, with new clusterings generated by randomly reassigning labels to $p$ proportion of objects outside the first cluster.}

\label{fig}
\end{figure}

In the first experiment, we follow \cite{graph_simulation_study} and begin with a ground truth labeling of 1,024 objects divided into 32 equally sized clusters. All objects are then randomly re-assigned to $c \in\{1, \ldots, 1024\}$ clusters to assess the constant baseline property and cluster count bias of the similarity measures. As documented in the literature \cite{vinh_first_nmi, newman_rmi, vinh_review, graph_simulation_study}, Figure \ref{fig}(a) shows that NMI exhibits a notable bias for higher cluster counts and clearly violates the constant baseline property. In contrast, the other four measures align closely with the desired behavior: the measured similarity between the ground truth and randomly generated clusterings remain near zero regardless of the number of clusters $c$.

Starting from the same ground truth specification, the second experiment randomly merges or splits clusters until there are exactly $c$ clusters in the newly created labeling. For values of $c<32$, two random clusters are repeatedly chosen and merged until there are $c$ clusters left. For $c > 32$, a randomly chosen cluster is split into two, with the size proportions of the resulting clusters drawn from $\text{Unif}(0,1)$. Intuitively, the measure of similarity between the ground truth and modified labelings should tend to $0$ as $c\to 1$ and $c \to 1024$, as the trivial clusterings that assign each object to an identical or unique label no longer contain any information about the ground truth. Figure \ref{fig}(b) shows that only AMI, ARI, and ResMI satisfy both properties. Notably, NMI's bias for higher cluster counts appears once again, with the measure remaining elevated even for values of $c \gg 32$.

Next, we consider the similarity between the same symmetric ground truth and a newly generated clustering with $p \in [0,1]$ proportion of  objects randomly shuffled between clusters. By a natural extension of the constant baseline property, the measure of similarity should  decrease monotonically in $p$, ranging from $1$ at $p=0$ to $0$ at $p=1$. As can be seen in Figure \ref{fig}(c), NMI and RMI are the only metrics that fail to satisfy this property---the former due to its bias for higher cluster counts, and the latter due to its correction term inducing negative values. 

Finally, we perform the same experiment, but with an asymmetric ground truth specification that assigns 50\% of the objects to one cluster and 25\% to another, while the remainder are evenly distributed across five additional clusters. When shuffling cluster assignments, the $p \in [0,1]$ proportion of objects that are \emph{not} in the first cluster are chosen, such that the 512 objects originally in the main cluster remain untouched. For instance, in the case of $p = 1$, $50\%$ of all objects (i.e. all objects not in the first cluster) are randomly assigned to new clusters. As seen in Figure \ref{fig}(d), ARI differs significantly from the other four measures, reflecting its poor performance in settings with an asymmetric distribution of cluster sizes, as identified previously in \cite{ARI_issue}. 

\begin{table}[htbp]
\caption{Properties of Clustering Similarity Measures}
\vspace{-27pt}
\begin{center}
$$ \begin{array}{c|ccccc}
 & \text{NMI} & \text{AMI} & \text{ARI} & \text{RMI} & \text { ResMI} \\
\hline \text{Constant Baseline} & \times & \checkmark  & \checkmark & \checkmark & \checkmark \\
\text{Model Independence} & \checkmark & \times & \times & \checkmark & \checkmark \\
\text{Constrained to $[0,1]$} &  \checkmark  & \times  & \times  &  \times  &  \checkmark \\
\text{Free of Cluster Count Bias} & \times & \checkmark & \checkmark & \checkmark  & \checkmark \\
\text{Free of Symmetry Bias}  & \checkmark & \checkmark & \times & \checkmark & \checkmark  
\end{array} $$
\label{table_checks}
\end{center}
\end{table}

Overall, ResMI appears to be relatively more conservative in terms of reporting lower similarity measures for generated clusterings that depart from the ground truth. In the second experiment, ResMI reaches its lower bound at the fastest rate as the number of clusters depart from the ground truth of $c= 32$, and the same occurs in the third experiment as the proportion of shuffled labels $p$ increases. Moreover, ResMI satisfies the constant baseline property, while being free of the cluster count and symmetry bias that NMI and ARI exhibit across the four experiments. Table \ref{table_checks}  summarizes the experimental results and properties of the five similarity measures we consider.

\section{Application: Contact Tracing Networks}

To further examine ResMI's performance as a measure of similarity between clusterings, we test its ability to identify meaningful communities in two real network datasets that contain information on the face-to-face interactions of workers in an office building \cite{contact_data_1, contact_data_2}. In both networks, each node represents an office worker, and each edge represents a recorded instance of a face-to-face interaction between the two corresponding office workers. Ground truth labels are given by each worker's department in the office building. 

\begin{figure}[htbp] 
\centering
\includegraphics[width=9cm]{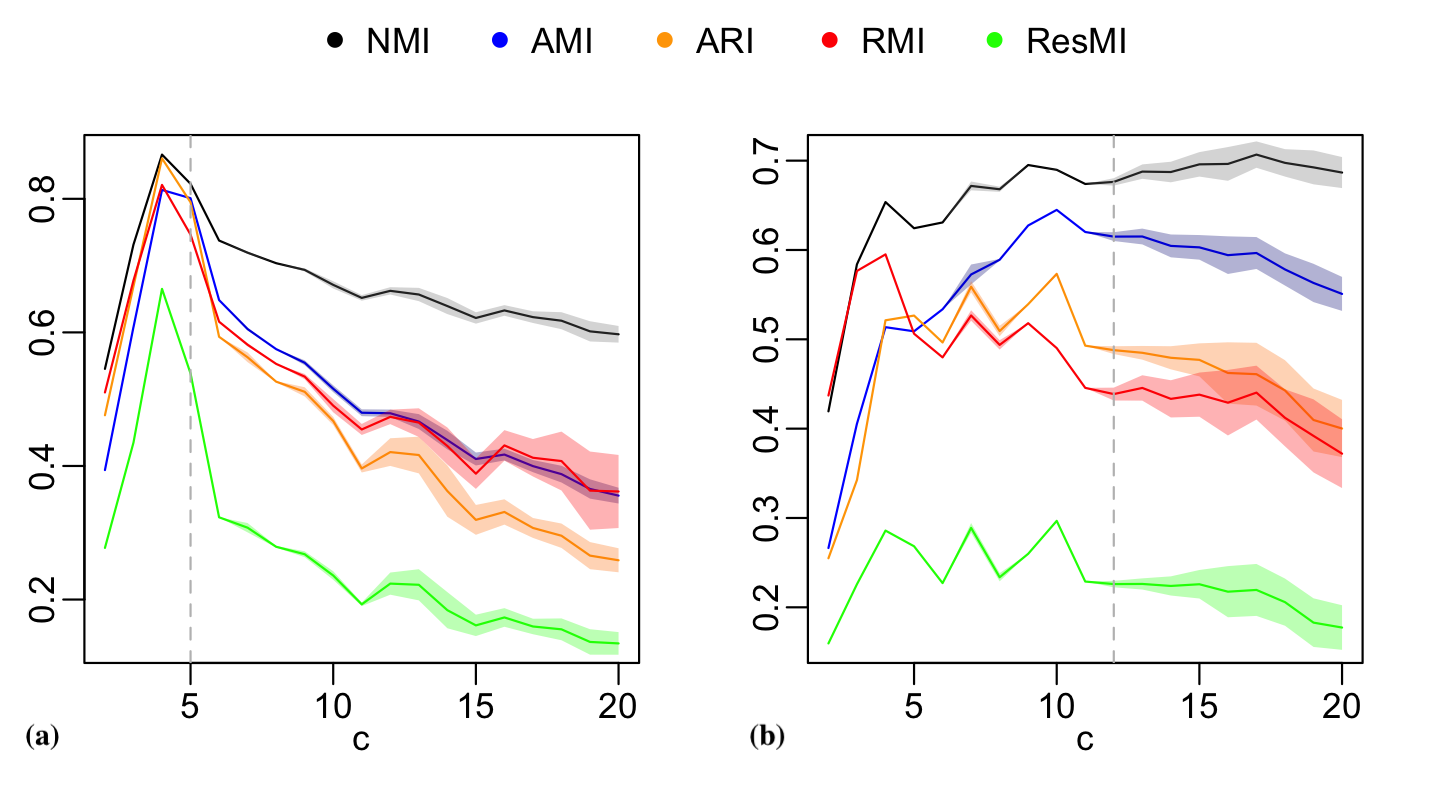}
\caption{Similarity between ground truth and estimated community labels by SCORE+ for varying values of $c$, averaged over 100 runs with error bars denoting one standard deviation. Dotted gray lines mark the ground truth number of communities. (a) Implementation on the contact tracing network of \cite{contact_data_1}. (b) Implementation on the contact tracing network of \cite{contact_data_2}. }
\label{fig3}
\end{figure}

To estimate community labels in  the two networks, we implement the SCORE+ algorithm of \cite{tracy_score_plus}, an extension of  SCORE \cite{score_original} that uses ratios of eigenvector entries of the graph adjacency matrix for spectral clustering. SCORE+ adds degree normalization to SCORE,  improving the procedure's performance in weak-signal settings. Like most spectral methods, SCORE+ takes the number of communities $c$ as a parameter. For the two contact tracing networks we consider, Figure \ref{fig3} shows the values of NMI, AMI, ARI, RMI, and ResMI between the ground truth and fitted community labels for varying values of $c$. 

\begin{figure}[htbp] 
\centering
\includegraphics[width=9cm]{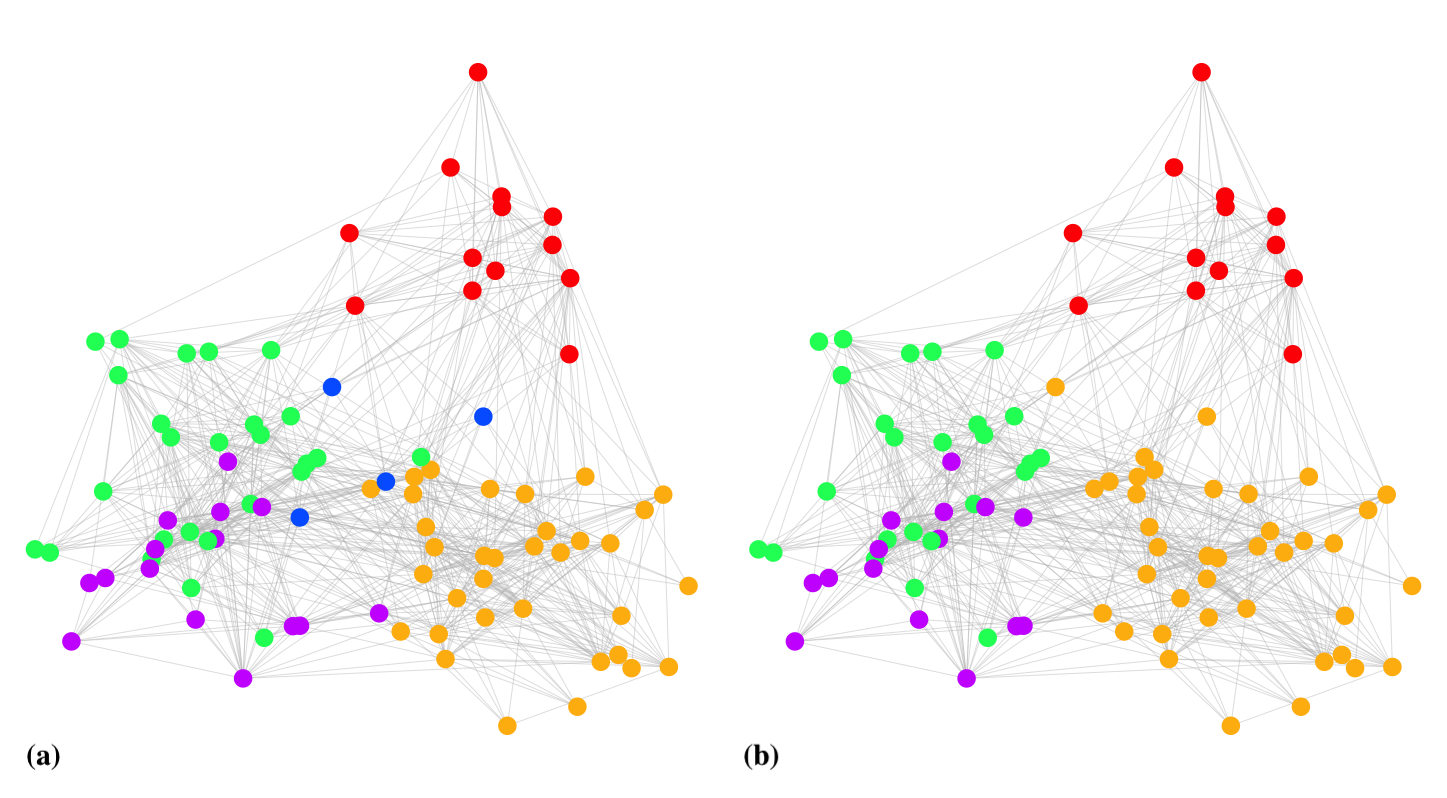}
\caption{(a) Ground truth and (b) estimated community labels (with $c=4$) for the contact tracing network of \cite{contact_data_1}. There are 92 nodes and 755 edges, and the network layout was determined by multidimensional scaling. The distribution of nodes across  ground truth communities is $\{ 34 , 26, 15 ,  13, 4\}$ and the distribution of nodes across estimated communities is $\{ 39 ,25, 15,  13\}$.}
\label{fig4}  
\end{figure}

Figure 2(a) shows that all five measures agree that the fitted community labels when $c=4$ is the most similar to the ground truth in the first contact tracing network, which has 5 different communities. Figure \ref{fig4} plots the ground truth community labels against the community labels estimated by SCORE+ when $c=4$. The two community labelings agree closely, with the only major discrepancy being the 4-member community near  the center, whose members are absorbed into neighboring communities in the estimated version.

\begin{figure}[htbp] 
\centering
\includegraphics[width=9cm]{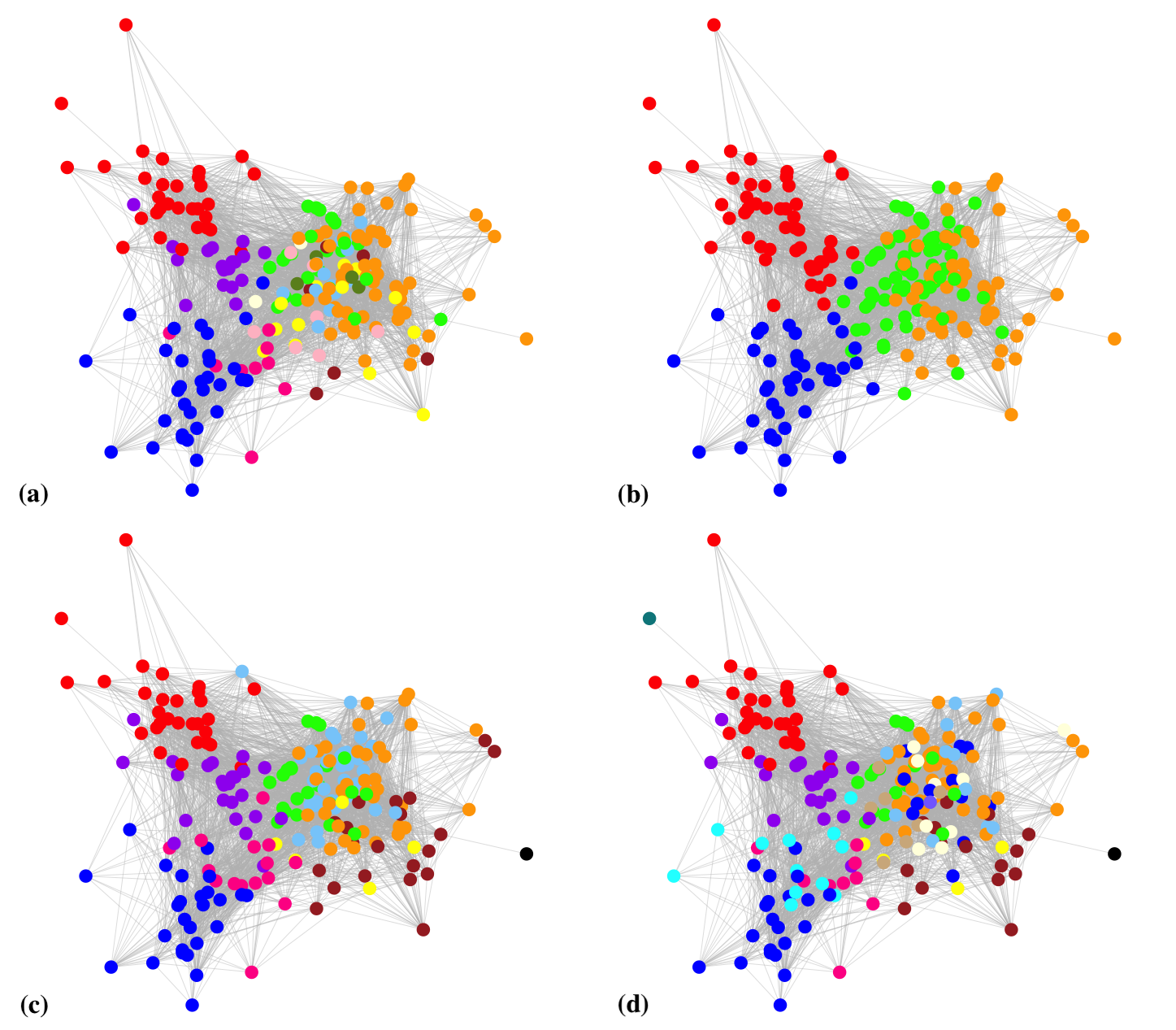}
\caption{(a) Ground truth community labels, and estimated community labels with (b) $c=4$, (c) $c=10$, and (d) $c=17$  for the contact tracing network of \cite{contact_data_2}. There are 217 nodes and 4,274 edges, and the network layout was determined by multidimensional scaling. The distribution of nodes across  ground truth communities is $\{  57 , 32  , 31 ,  23 ,  18  , 14 ,  13  ,  9   , 7 ,   7 ,    4   , 2 \}$. The distribution of nodes across the estimated communities are (b) $\{69, 58, 49 ,41\}$, (c) $\{ 37 ,37,29, 26, 24, 22 , 21,  15 ,  5,  1  \}$, and (d) $\{ 33, 25 ,23, 22 ,17 ,17 ,16, 13 ,10 ,10, 10,  8 , 6 , 4 , 1,  1 , 1 \}$.}

\label{fig5}
\end{figure}

On the other hand, the optimal number of communities in the second contact tracing network differs significantly across the five similarity measures. While there are 12 ground truth communities, Figure \ref{fig3}(b) shows that the estimated community labels are most similar to the ground truth when $c=4$ according to RMI, $c=10$ according to AMI, ARI, and ResMI, and $c=17$ according to NMI. Figure \ref{fig5} shows the contact tracing network with the ground truth and estimated community labels for $c \in \{4, 10, 17\}$.

Figures \ref{fig5}(a) and \ref{fig5}(b) reveals that the estimated community structure is overly simplistic for $c = 4 $, as it fails to capture overlapping communities near the center of the plotted network. This suggests RMI may be too conservative, particularly in settings where there are a higher number of ground truth communities. Conversely, Figure \ref{fig5}(d) shows that the estimated community structure is excessively fragmented when $c = 17$, with numerous small communities replacing the larger ones in the ground truth labeling.  This finding aligns with experimental results in the previous section, which established NMI's tendency to favor clusterings with a higher number of clusters. As displayed in Figure \ref{fig5}(c), the estimated community labels when $c= 10$ strikes the best balance between simplicity and complexity, correctly identifying the majority of large and medium-sized ground truth communities across the network. Given that this labeling was deemed most similar to the ground truth by AMI, ARI, and ResMI, our results further highlight the potential value of ResMI as a measure of clustering similarity.

\section{Conclusion}

This paper introduced ResMI, a novel clustering similarity measure that integrates principles from information theory and pair counting methods. Our experiments on synthetic and real-world datasets demonstrate that ResMI satisfies the constant baseline property and remains robust against the biases of existing similarity measures. Defined entirely in information theoretic terms without adjustments for chance correction, ResMI is both model-independent and intuitively interpretable, enhancing its utility as a clustering similarity measure.

Future research could investigate ResMI’s scalability to larger datasets with complex community structures and examine potential limitations in these settings. Additionally, a promising direction would be to incorporate ResMI directly into clustering and community detection algorithms.

\section*{Acknowledgment}

We thank Flavio du Pin Calmon and Sajani Vithana for constructive comments and discussions.

\bibliographystyle{IEEEtran}


\end{document}